\documentclass[12pt,preprint,tighten]{aastex}
\usepackage[]{times, graphicx}
\newcommand{\pref}{\protect\ref}
\newcommand{\soho}{{\em SOHO{}}}
\newcommand{\trace}{{\em TRACE{}}}
\newcommand{\hinode}{{\em Hinode{}}}

\begin{document}

\shorttitle{Roots of Coronal Heating}
\shortauthors{De Pontieu, McIntosh, Hansteen \& Schrijver} 
\title{Observing the Roots of Solar Coronal Heating - in the Chromosphere} 
\author{Bart De Pontieu\altaffilmark{1,4}, Scott W. McIntosh\altaffilmark{2}, Viggo H. Hansteen\altaffilmark{3,1}, Carolus J. Schrijver\altaffilmark{1}} 
\altaffiltext{1}{Lockheed Martin Solar and Astrophysics Lab, 3251 Hanover St., Org. ADBS, Bldg. 252, Palo Alto, CA 94304} 
\altaffiltext{2}{High Altitude Observatory, National Center for Atmospheric Research,P.O. Box 3000, Boulder, CO 80307}
\altaffiltext{3}{Institute of Theoretical Astrophysics, University of Oslo, Blindern, Oslo 315, Norway}
\altaffiltext{4}{Electronic address: bdp@lmsal.com}

\begin{abstract}
The Sun's corona is millions of degrees hotter than its 5,000~K photosphere. This heating enigma is typically addressed by invoking the deposition at coronal heights of non-thermal energy generated by the interplay between convection and magnetic field near the photosphere. However, it remains unclear how and where coronal heating occurs and how the corona is filled with hot plasma. We show that energy deposition at coronal heights cannot be the only source of coronal heating, by revealing a significant coronal mass supply mechanism that is driven from below, in the chromosphere. We quantify the asymmetry of spectral lines observed with \hinode{} and \soho{} and identify faint but ubiquitous upflows with velocities that are similar (50-100 km/s) across a wide range of magnetic field configurations and for temperatures from 100,000 to several million degrees.  These upflows are spatio-temporally correlated with and have similar upward velocities as recently discovered, cool (10,000 K) chromospheric jets or (type II) spicules. We find these upflows to be pervasive and universal. Order of magnitude estimates constrained by conservation of mass and observed emission measures indicate that the mass supplied by these spicules can play a significant role in supplying the corona with hot plasma. The properties of these events are incompatible with coronal loop models that only include nanoflares at coronal heights. Our results suggest that a significant part of the heating and energizing of the corona occurs at chromospheric heights, in association with chromospheric jets.
\end{abstract}



\section{Introduction}
Statistical correlations between the brightness of the chromosphere and corona strongly suggest that these two regions are powered by a similar driver \citep{Schrijver2000}. Yet, despite some suggestions that much of the non-thermal energy driving coronal heating is deposited in the chromosphere \citep{Aschwanden2006,Gudiksen2005,McIntosh2007a}, most models of coronal loops treat the chromosphere as a passive mass reservoir that merely reacts to energy release at coronal heights and the associated changes in the coronal thermal conductive flux or deposition of non-thermal energy \citep{Klimchuk2006}. The differing role of the chromosphere in these models implies that observations of how the corona is filled with plasma from the chromosphere hold significant diagnostic potential for distinguishing between various coronal heating models \citep{Klimchuk2006, Patsourakos2006}. Yet, it has been very difficult to establish how the dynamics and energetics of chromosphere and corona are coupled \citep{Klimchuk2006, McIntosh2007b,Aiouaz2007,Hansteen2007}.
 
Chromospheric spicules have long been considered a candidate for such coupling by providing mass to the corona and/or solar wind \citep{Beckers1968,Pneuman1978,Athay1982}. These jet-like features, with lifetimes of order 3-10 minutes, propel cool matter upwards to coronal heights with speeds of 20-30 km/s \citep{Beckers1968}. A significant fraction of these are likely caused by magnetoacoustic shocks on magnetic flux concentrations \citep{DePontieu2004,Hansteen2006,DePontieu2007a}. Spicules are observed at temperatures from 5,000 to 500,000 K and are estimated to carry upward a mass flux 100 times larger than that of the solar wind \citep{Pneuman1978}. However, no signatures of these ejecta at coronal temperatures have been reported previously \citep{Withbroe1983,Mariska1992}. Consequently, a direct role of spicules in the coronal mass/energy balance has been dismissed as unlikely \citep{Withbroe1983}. 

However, the high spatio-temporal resolution of the Solar Optical Telescope \citep[SOT;][]{Tsuneta2008} onboard \hinode{} \citep{Hinode} has revealed a previously unrecognized second type of spicules \citep{DePontieu2007b} that is a prime candidate for establishing a link between corona and chromosphere. These ``type II'' spicules have much shorter lifetimes (10-100~s), are more violent (upflows of order 50-150~km/s), and often fade rapidly from the chromospheric \ion{Ca}{2} H 3968\AA{} SOT passband, which suggests rapid heating and ionization of singly to at least doubly ionized calcium. The formation of these ubiquitous type II spicules has been attributed to magnetic reconnection \citep{DePontieu2007b}.

To study the thermal evolution of these spicules, we use simultaneous observations of \ion{Ca}{2} H emission (with SOT) and EUV spectral lines formed in the corona at temperatures of several million K from the Extreme-ultraviolet Imaging Spectrometer \citep[EIS;][]{Culhane2007}. We use spectra from SUMER \citep[][]{Wilhelm1995} on SOHO \citep[][]{Fleck1995} to show that the process we describe is ubiquitous and occurs in active regions, quiet Sun and coronal holes. The observations and our novel analysis techniques are described in \S~\pref{robs}, while \S~\pref{obs} covers the results of our analysis. In \S~\pref{discuss} we discuss the impact of our results on the existing coronal heating hypotheses.

\section{Observations \& Analysis}\label{robs}

We focus on plage regions in active regions NOAA 10976 and 10977 observed near disk center on 2 and 5 December 2007 by Hinode/SOT and EIS. By observing at disk center we reduce the ambiguities introduced by line-of-sight superposition at the limb (where spicules are most readily identified), which has plagued such studies in the past \citep{Withbroe1983}. SOT obtained a timeseries of images using the \ion{Ca}{2} H 3968 \AA{} filter (with 0.5s exposure and 8s cadence) that covered a 25$\arcsec\times$40$\arcsec$ area of plage. Simultaneous EIS observations of the same region started with a region-wide scan with coarse horizontal (5$\arcsec$) stepping with 60s exposures for a wide variety of spectral lines. This was followed by much finer (1$\arcsec$ stepping), rapid (60s exposure) coronal rasters with EIS over a 5$\arcsec$x80$\arcsec$ sliver of plage. 

To connect the chromospheric and coronal dynamics we need to coalign the coronal spectral rasters with the high cadence chromospheric imaging. This is not straightforward. The SOT timeseries are coaligned onboard \hinode{} using a correlation tracker. The EIS spectra are pointed independently from SOT, and undergo significant jitter caused by the spacecraft, and thermal flexing within EIS. We remove the spacecraft jitter from the EIS pointing by using the xrt\_jitter IDL routine (from solarsoft). To correct for the effect of thermal flexing on EIS pointing, we use several steps. First, we perform additional coalignment of the \ion{Ca}{2} timeseries using cross-correlation. This removes drifts caused by solar evolution within the small field of view of the SOT correlation tracker. We use chromospheric and coronal images from \trace{} \citep[][]{Handy1999} to connect the EIS and SOT data and determine the EIS pointing as a function of time.

This pointing information is used to calculate synthetic rasters of upper chromospheric activity (UCA) derived from the \ion{Ca}{2}H timeseries. The latter is dominated by slowly evolving (on timescales $>60$s) photospheric and low-chromospheric contributions because of the wide bandpass (FWHM=2.2\AA) of the SOT \ion{Ca}{2} H filter. To isolate the relatively faint signature (from the Ca H core) of the upper-chromospheric type-II spicules in the \ion{Ca}{2} H timeseries, we exploit their highly dynamic nature: they evolve on much shorter timescales ($<60$s) \citep[][]{DePontieu2007b} than the photospheric and low-chromospheric contributions. We thus perform, for each pixel, temporal high-pass Fourier filtering on the timeseries with a cutoff of 18 mHz. The resulting timeseries of (the absolute value of the) high frequency signal is a good proxy for UCA because it removes the photospheric and low-chromospheric features (online movie 1).

To study the high temperature response, we analyze EIS spectra of moss regions which constitute the transition region (TR) at the footpoints of active region loops \citep{Berger1999,DePontieu1999,Fletcher1999,Martens2000}. We find that the bright core of the coronal spectral lines are typically Doppler shifted by less than 10-20 km/s and note that the Doppler shift and intensities of the coronal profiles show no spatial correlation between upper-chromospheric and coronal activity \citep{Hansteen2007}. However, \citet{Hara2008} recently discovered that there is often a faint blue-shifted component in addition to the bright core of the line, indicating coronal upflows of order 50-100 km/s. To isolate these upflows from the bright core of the spectral line more clearly, we calculate maps of the blue-red (B-R) asymmetry in the coronal spectral lines (Fig.~\ref{fig1}) and find that the upflows occur preferentially in moss footpoint regions.

To calculate the B-R asymmetry of spectral lines, we first fit a single Gaussian to the line profile $I_\lambda$, and determine the line centroid $\lambda_0$ of the Gaussian fit. The B-R asymmetry for an offset $\Delta\lambda_1$ from the line centroid is then given by:
\begin{equation}
BR_{\Delta\lambda_1}=\sum^{\lambda_0-\Delta\lambda_1+\delta\lambda_w}_{\lambda_0-\Delta\lambda_1-\delta\lambda_w}I_{\lambda}-\sum^{\lambda_0+\Delta\lambda_1+\delta\lambda_w}_{\lambda_0+\Delta\lambda_1-\delta\lambda_w}I_{\lambda}
\end{equation}
with $\delta\lambda_w$ the wavelength range over which the B-R asymmetry is determined. To determine the propagated errors on the B-R asymmetry measure, we use estimates for the error on the intensity using the EIS software (including photon noise and other errors). In our maps we only show locations where the B-R asymmetry is larger than twice the estimated propagated error.
We apply the same method to SUMER rasters of quiet Sun and coronal hole on 6-nov-1999, after calibrating the data using the techniques described by \citet{Davey2006}.

\section{Results}\label{obs}

We calculate B-R asymmetry maps for EIS rasters of \ion{Fe}{14} 274\AA{} (formed at 2 MK under equilibrium conditions) for velocities arond 80 km/s and 120 km/s (Fig.~\ref{fig2}b,c). Both B-R asymmetry maps show no obvious correlation with the intensity of the core of the \ion{Fe}{14} line (Fig.~\ref{fig2}a) from which they are derived. In contrast, we find that the sites of faint upflows at coronal temperatures (Fig.~\ref{fig2}b,c) are well correlated with upper-chromospheric activity (Fig.~\ref{fig2}d, online movie 2). The correspondence between the B-R and UCA maps is surprisingly good, given several complicating factors. 

First, despite our careful coalignment, there remains uncertainty of order 2\arcsec~in the EIS/SOT coalignment. In addition, the correlation between the chromospheric signal and coronal blueshifts associated with spicules depends greatly on the viewing angle between the line-of-sight and the direction of the upward flows (Fig.~5). When looking down at loop footpoints, we observe the chromospheric jet as a \ion{Ca}{2} H brightening (because of projection), associated with a faint, strongly blue-shifted component in the coronal line. When the line-of-sight is not aligned with the upflows, the spicules will appear as jets in the chromospheric line, but we no longer see blue-shifts in the coronal line since the line-of-sight velocity is reduced, and the spicular emission measure is too small compared to the dominant coronal emission. The wide range of angles between upflows (i.e., magnetic field) and line-of-sight thus naturally leads to an imperfect correlation between chromospheric brightness and coronal upflows. It is also possible that some of the UCA sites that do not match coronal upflow events do have coronal counter parts, but at much higher or lower temperatures than the \ion{Fe}{14} line we observe. We also cannot expect a one-to-one correlation between the upper chromospheric brightness and the brightness in the coronal asymmetry maps, because we do not know the exact formation and heating mechanism of the spicules. Finally, some of the coronal upflow events could be nanoflares that are driven by energy deposition in the corona, i.e., without significant chromospheric brightness enhancement. 

The relationship between these chromospheric events and coronal upflows is elucidated by our finding that the velocities of the latter in two \ion{Fe}{14} lines (Fig.~\ref{fig3}) are similar to the upward velocities for type II spicules seen at the limb \citep[from][]{DePontieu2007b}, both peaking between 50 and 100km/s. We find that these upflows are pervasive and universal: UV lines observed with SUMER show faint upflowing components with velocities of the same magnitude (50-100km/s) that are well correlated across the TR and low corona (100,000 to 600,000 K) for a whole range of magnetic field configurations (active region plage, and network in quiet Sun and coronal holes, with the velocities in the latter two somewhat lower than in plage). 

This is illustrated with spectroheliograms of \ion{C}{4} 1548\AA{} and \ion{Ne}{8} 770\AA{} from for a coronal hole region near disk center (Fig.~\ref{fig4}). We find a clear difference in \ion{Ne}{8} intensity (reduced) and Doppler velocity (blueward) as well as \ion{C}{4} Doppler velocity (blueward) in the coronal hole compared to the quiet Sun. These differences can be understood in terms of a picture where the bulk of the material reaching the formation temperature of \ion{Ne}{8} is injected into the solar wind \citep{McIntosh2007b}. We also find that the average profile in the network in both quiet Sun and coronal hole shows significant asymmetry towards the blue for velocities of 50-100 km/s for both \ion{C}{4} and \ion{Ne}{8} lines. The B-R asymmetry maps show that faint upflows at 50-100 km/s are well correlated for a lower TR (\ion{C}{4}) and a coronal (\ion{Ne}{8}) line and preferentially occur in and around the network. The spatial patterning and correlation of these upflows does not change as one crosses the boundary between quiet Sun and coronal hole. This is surprising, since the magnetic field configuration in the corona is dramatically different.

\section{Discussion}\label{discuss}

Our analysis of the asymmetry of spectral lines is very sensitive to the presence of blends in the wings of the lines, which can skew the asymmetry maps towards the red or blue so that these are no longer indicative of flows. For that reason, we have avoided lines that have strong blends. We have extensively studied the influence of blends on the lines used, and find that none of them change our conclusions (see a follow-up paper for a detailed discussion). We point out that the upflow maps for different \ion{Fe}{14} lines (at 264 and 274\AA) are very similar, and that we observe similar upflows for a wide range of temperatures (\ion{C}{4} 1548\AA{}, \ion{Ne}{8} 770\AA{}, \ion{Fe}{14} 264 and 274\AA{}). 

\begin{figure}
\epsscale{1}
\plotone{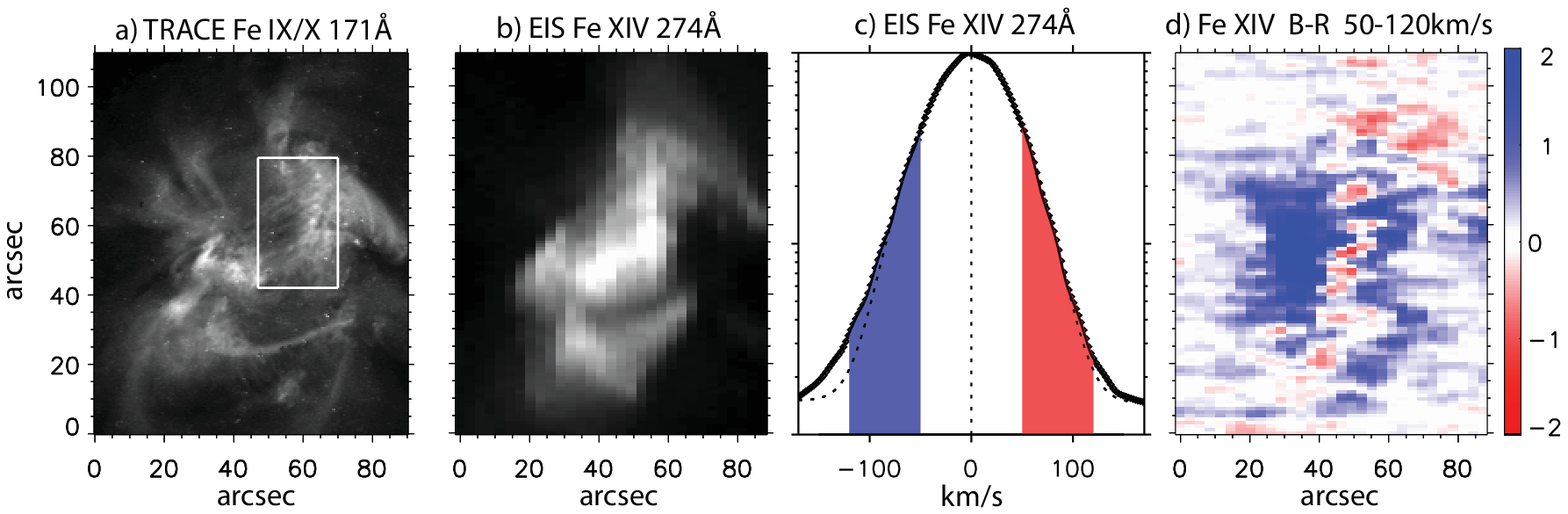}
\caption{Fe~IX/X 171\AA{} images (from \trace) and EIS rasters in \ion{Fe}{14} 274\AA{} of NOAA AR 10977 on 5-dec-2007 show cooler moss footpoints (log T=5.9, a) as well as hotter loops (log T= 6.2, b) connecting the footpoints. Spectral line profiles of the coronal \ion{Fe}{14} line show a deviation from a Gaussian (dashed line, c) indicative of hot plasma flowing upwards at high speeds. Spatial maps of the B-R asymmetry at 50-120 km/s (d, see text for details) show a good correspondence of blueshifted events with the locations of moss footpoint regions (a). Online movie 1 of the region indicated by the white box in panel (a) shows how high-pass Fourier filtering isolates the upper-chromospheric activity (right panel of movie) from the more slowly evolving photospheric signals that dominate the Hinode \ion{Ca}{2} H bandpass (left panel of movie).
\label{fig1}}
\end{figure}

\begin{figure*}
\epsscale{0.5}
\plotone{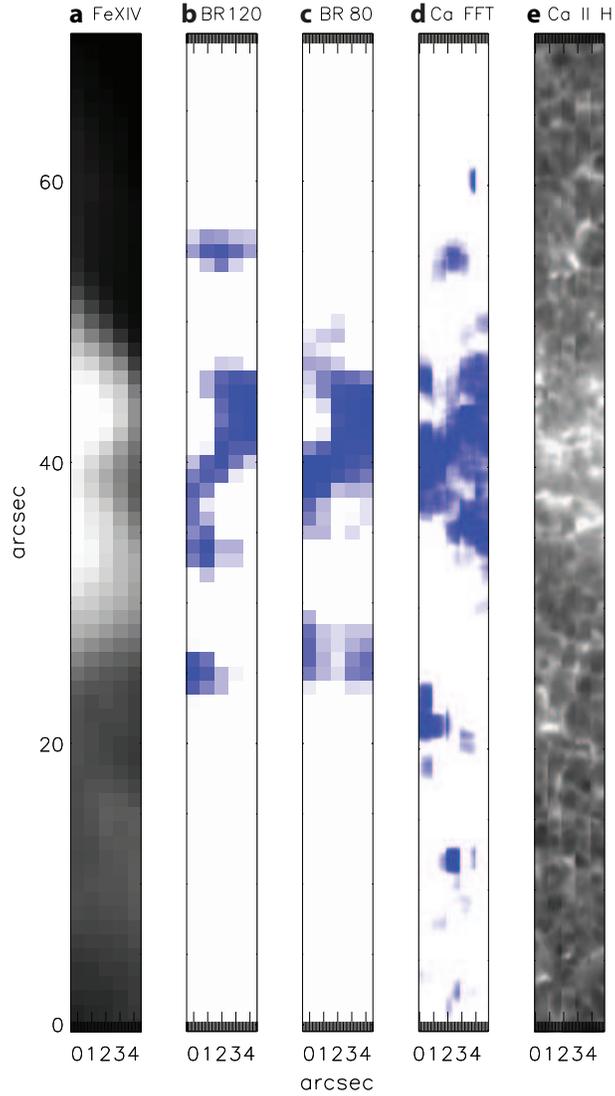}
\caption{The locations of blueshifted signals from the B-R asymmetry maps of \ion{Fe}{14} 274\AA{} (panel a shows the intensity) centered around 120 (panel b) and 80km/s (panel c) correlate well with a synthetic raster map of the upper chromospheric activity (panel d, which is derived from panel e) of a small plage region on 2-dec-2007. Online movie 2 shows a timeseries of similar comparisons. \label{fig2}}
\end{figure*}

\begin{figure*}
\epsscale{0.6}
\plotone{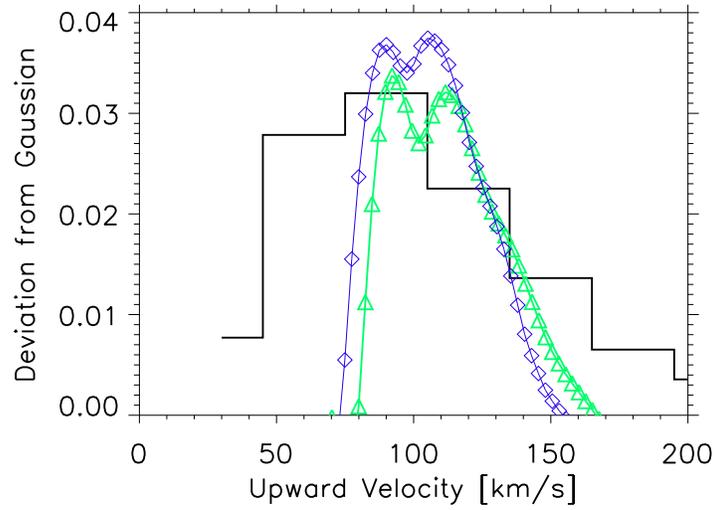}
\caption{A histogram of upward velocities in chromospheric type II spicules (black line) as derived from \ion{Ca}{2} H 3968\AA{} timeseries at the limb \citep[from][]{DePontieu2007b} shows a range of velocities that is similar to coronal upflow velocities derived from subtracting a Gaussian fit from the profile of \ion{Fe}{14} 264\AA{} (blue) and \ion{Fe}{14} 274\AA{} (green)
lines for all locations that show significant upflows in active region 10977 on 5-dec-2007. \label{fig3}}
\end{figure*}

\begin{figure}
\epsscale{1}
\plotone{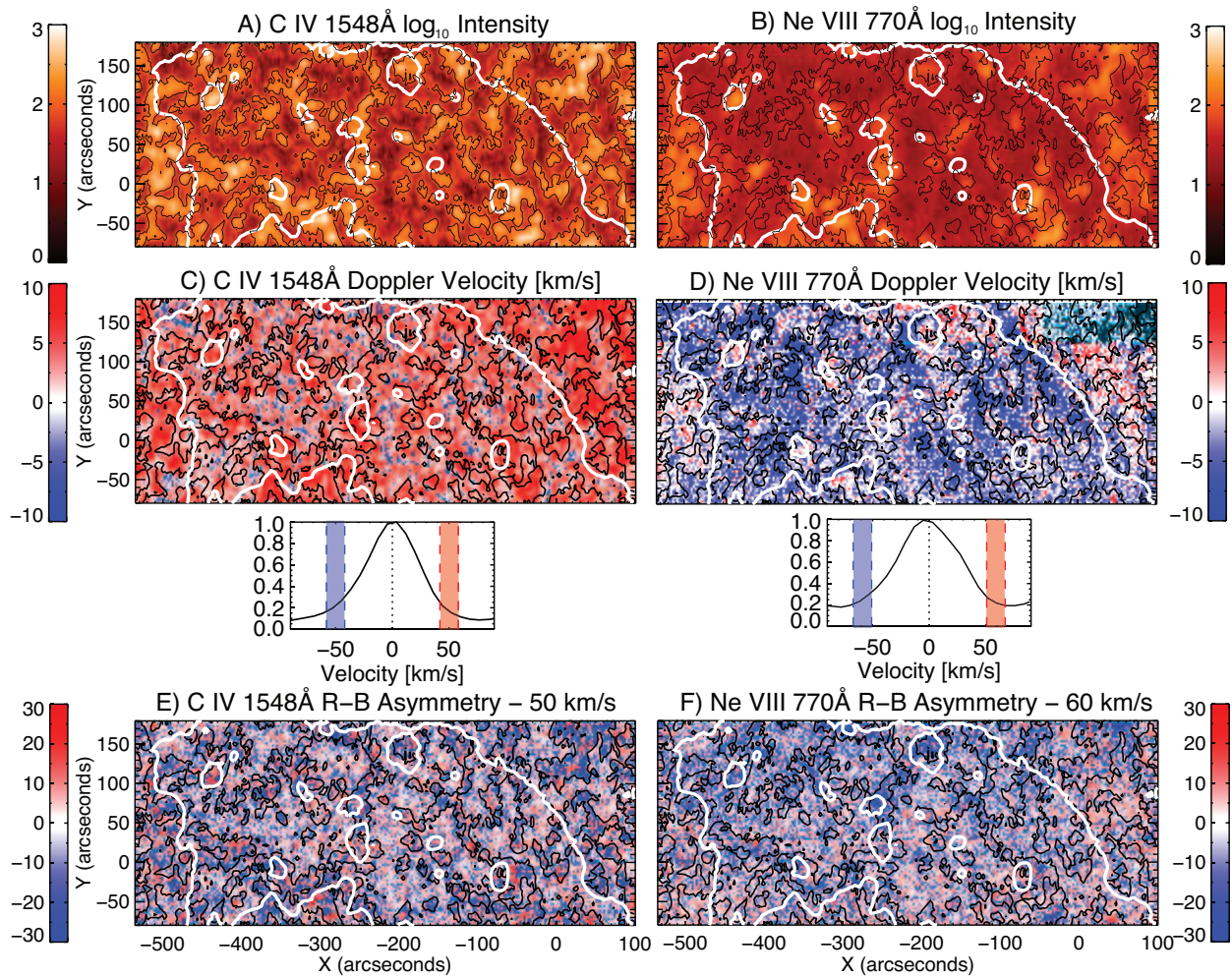}
\caption{Panels a and b show the logarithm of intensities of \ion{C}{4} 1548\AA\ and \ion{Ne}{8} 770\AA\ for an equatorial hole observed at disk center with SUMER. The solid contours are a proxy for the supergranular network boundaries (derived from a threshold of 150 counts in panel a). The solid white contour delineates the boundary of the coronal hole from the quiet sun. Panels c and d show the Doppler shift of the lines. Below this we show representative normalized network line profiles taken from x=-300",y=-42". We determine B-R asymmetry maps (panels e and f) for a 24 km/s wide region [shown as red/blue rectangles] at offsets from line center of 50km/s for \ion{C}{4} and 60km/s for \ion{Ne}{8}. In both cases, we see strong enhancement of the blue wing emission in the magnetic network.  \label{fig4}}
\end{figure}

We estimate that these spicule-associated upflows play a significant role in replenishing the corona with hot mass. To play such a role in active regions, these hot upflows must provide enough mass flux $f_i=\rho_i v_i N_i$ to replenish the corona which loses mass by downflows (after cooling) at a rate of $f_c=\gamma\rho_c h_c/t_c$. Here, $\rho_i$ is the density of the coronal mass propelled upward and heated in a spicule, $v_i$ the upward velocity of the spicule, $N_i$ the number of spicules that occur at any location over the spicule lifetime $t_i$, $\rho_c$ the density in coronal loops, $h_c$ the coronal scale height, $t_c$ the time for coronal plasma to cool to chromospheric temperatures, and $\gamma$ the ratio of cross sections of the upper corona and the plage region below. Our observations in coronal lines also show that the faint blue-shifted spicular emission measure $e_i=(\epsilon_T\rho_i)^2v_it_iN_i$  (with $\epsilon_T$ the fraction of total spicular coronal density at a temperature $T$) can only be a fraction $\alpha$ of that of the dominant emission $e_c=\rho_c^2h_c$ at the core of the lines, which is emitted by the slowly cooling plasma in previously filled loops. We thus require that $f_i=f_c$ and $e_i=\alpha e_c$, and find that $N_i=\gamma^2\epsilon_T^2 h_ct_i/(\alpha v_it_c^2)$, and $\rho_i=\alpha\rho_ct_c/(\gamma\epsilon_T^2t_i)$. For realistic estimates of $\gamma=3$, $h_c=50,000$km, $t_c=2000$s \citep{Schrijver2000}, $\alpha=0.05$ (from analysis of EIS spectra, Fig.~\ref{fig1}c,d), $t_i=100$s and $v_i=100$km/s \citep{DePontieu2007b} we find $\rho_i\approx\rho_c/(3\epsilon_T^2)$ and $N_i\approx 2\epsilon_T^2$. \ion{Fe}{14} observations only cover temperatures around 2 MK, and most likely reveal only a fraction of the coronal component of spicules, which may range in temperature from 1 to 10MK. If we make the reasonable assumption that only a fraction $\epsilon_T=0.25\--0.5$ of the spicular coronal mass flux is observed in \ion{Fe}{14}, we find, for coronal densities $\rho_c\sim10^9$cm$^{-3}$, a total density of the coronal component of spicules of order $1-5\times10^9$cm$^{-3}$. 

This implies that these spicules can fill the corona with hot plasma even if only between 1 and 5 \% of the chromospheric spicule densities reach coronal temperatures \citep{Beckers1968,Pneuman1978}. We can also estimate how many spicules $N_p$ are required at any time in a plage region of diameter $\delta$: $N_p=N_i\delta^2/\delta_i^2$  with $\delta_i$ the diameter of a spicule. For $\delta\sim8000$km and $\delta_i\sim250$km, $N_p$ is of order 80\--320, implying a spicule number density of 0.7\--3 spicules per square arcsecond of plage. This density is compatible with SOT observations, which implies that these spicules can play a significant role in filling coronal loops. Individual spicular events have energies of the order of nanoflares ($\sim10^{23}$ erg) and must recur every $t_i/N_i$ seconds at the same location (i.e, 250 \--1000 s for the above estimates) to replenish the coronal mass. The energy flux carried into the corona by these hot spicules is of order $5\times10^6$ erg cm$^{-2}$ s$^{-1}$ with the above assumptions and if we assume that all spicules reach 2 MK temperatures. This is of the same order of magnitude as the energy flux required to heat the active region corona \citep{Priest1982}.

The spatio-temporal correlation between upper chromospheric activity (associated with jets) and faint upflows at several MK, and the ubiquity of faint and correlated upflows with velocities of order 50-100 km/s for temperatures ranging from 10,000 to 2 MK for a variety of magnetic field configurations (active region, quiet Sun and coronal hole) poses significant challenges to our current understanding of the mass cycling of the corona. The properties of these events do not seem compatible with those predicted by coronal loop models that are driven only by nanoflares at coronal heights. Such models predict high-speed upflows at very high temperatures of several million degrees, with either lower velocities or no detectable upflows at lower temperatures ($<$1 million K) \citep{Patsourakos2006}. More generally, our observations of similar upflow velocities for a wide range of temperatures from 10,000 K to 2 million K are a challenge to any model in which upflows are driven by transient overpressure at chromospheric heights \citep[e.g., by deposition of non-thermal energy from electron beams generated in the corona;][]{Raftery2009} because those lead to increasing velocities with temperature. 

Our finding that these upflows do not change qualitatively or quantitatively between quiet Sun and coronal holes strongly suggests that they are driven from below. This is because models in which energy is deposited at coronal heights are dependent on conductive flux and/or reconnection at tangential discontinuities, which are expected to have significantly different properties in open versus closed field environments \citep{Parker1989}. Such models also do not provide a natural explanation for why the chromospheric counterparts show upward extrusions and can be seen in lines from neutral elements such as Mg I b 5172 \AA{} (SOT). 

\begin{figure}
\epsscale{1}
\plotone{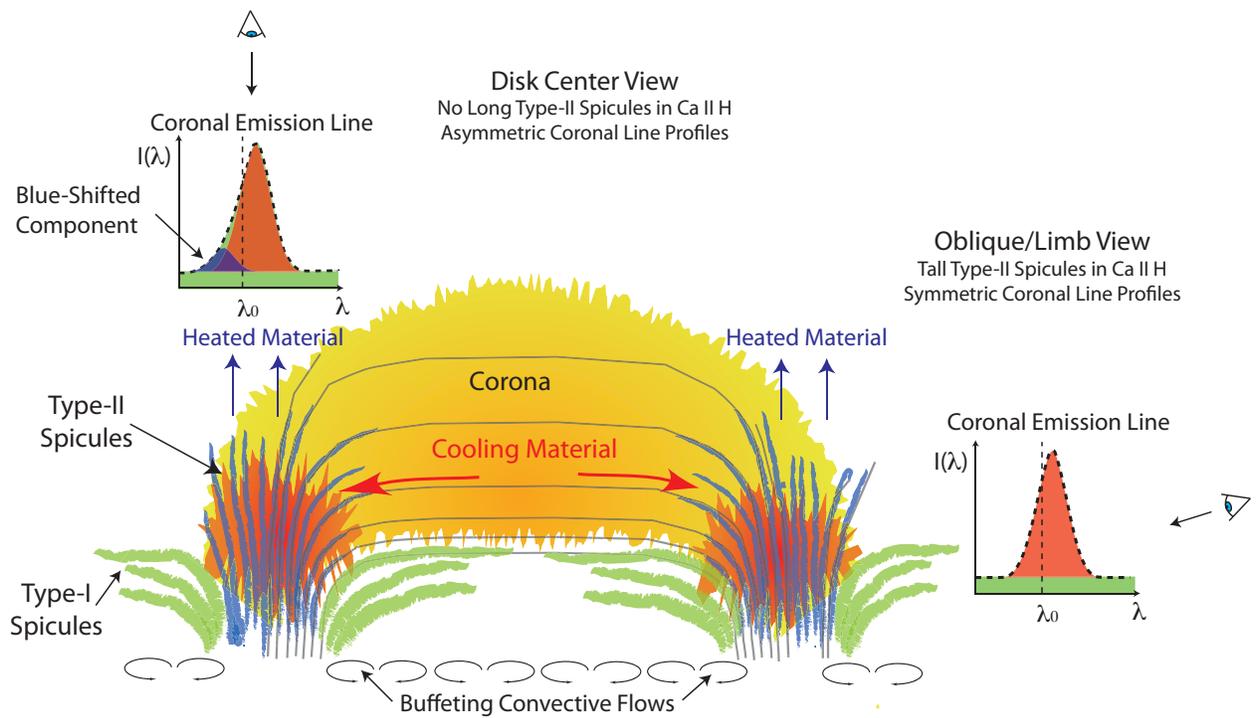}
\caption{Cartoon illustrating the mass and energy transport between the chromosphere, TR and corona, as deduced from SOT and EIS observations. See \S 4 for details. 
\label{fig5}}
\end{figure}

Nanoflares at coronal heights may be important for coronal heating, but they cannot explain our observations. Our analysis suggests a scenario (Fig.~\ref{fig5}) where a significant, potentially dominant, part of the heating of the coronal plasma occurs at chromospheric heights, in association with chromospheric jets \citep[similar to previous suggestions, e.g.,][]{Pneuman1978,Athay1982}. A significant fraction of the plasma propelled upward in these chromospheric jets or type II spicules is heated to coronal temperatures, providing the corona with hot plasma. The coronal emission of individual spicules is faint compared to the dominant emission, which originates in previously filled loops that are slowly cooling. Previous efforts to establish a connection between chromosphere and corona have focused on properties of the core of coronal lines. The weak emission of the coronal upflows associated with spicules, which provide the direct link to the chromosphere, helps explain why these efforts have failed in the past \citep[e.g.,][]{Hansteen2007}. 

Conceptually, our findings are compatible with results from numerical simulations \citep{Gudiksen2005,Hansteen2007b} that suggest that most of the heating in the solar atmosphere occurs below heights of 2 Mm, i.e., at chromospheric heights. Preliminary results of more recent simulations suggest a scenario in which magnetic reconnection in the chromosphere causes upward jets of plasma at a range of temperatures owing to magnetic Lorentz forces even as the plasma is heated to TR and coronal temperatures in the process.

\acknowledgements
SWM acknowledges support from NSF grant ATM-0925177 and NASA grants NNG06GC89G and NNX08AH45G. BDP is supported by NASA grants NNM07AA01C (HINODE), NNG06GG79G and NNX08AH45G. \hinode{} is a Japanese mission developed and launched by ISAS/JAXA, with NAOJ as a domestic partner and NASA and STFC (UK) as international partners.



\end{document}